\documentclass{IEEEtran}
\usepackage{wrapfig}
\usepackage{graphicx}
\usepackage[export]{adjustbox}
\usepackage{mdframed}
\usepackage{amsmath} 
\usepackage{amsfonts} 
\usepackage{amssymb} 
\usepackage[backend=biber,style=numeric,sorting=none]{biblatex} 
\title{Is Research Software Science a Metascience?}
\author{Evan Eisinger, Michael A. Heroux}
\date{} %
\begin{document}

\maketitle
\pagestyle{plain} 

\begin{abstract}
As research increasingly relies on computational methods, the reliability of scientific results depends on the quality, reproducibility, and transparency of research software. Ensuring these qualities is critical for scientific integrity and discovery. This paper asks whether Research Software Science (RSS)--the empirical study of how research software is developed and used--should be considered a form of metascience, the science of science. Classification matters because it could affect recognition, funding, and integration of RSS into research improvement. We define metascience and RSS, compare their principles and objectives, and examine their overlaps. Arguments for classification highlight shared commitments to reproducibility, transparency, and empirical study of research processes. Arguments against portraying RSS as a specialized domain focused on a tool rather than the broader scientific enterprise. Our analysis finds RSS advances core goals of metascience, especially in computational reproducibility, and bridges technical, social, and cognitive aspects of research. Its classification depends on whether one adopts a broad definition of metascience--any empirical effort to improve science--or a narrow one focused on systemic and epistemological structures. We argue RSS is best understood as a distinct interdisciplinary domain that aligns with, and in some definitions fits within, metascience. Recognizing it as such can strengthen its role in improving reliability, justify funding, and elevate software development in research institutions. Regardless of classification, applying scientific rigor to research software ensures the tools of discovery meet the standards of the discoveries themselves.
\end{abstract}

\section{Introduction}

Scientific inquiry today is upheld by an ever-growing reliance on computational methods. This fundamental change in the way research is conducted requires increased awareness of the need for robust, transparent, and reproducible research practices in all scientific disciplines~\cite{1}. The increasing dependence on computational methods represents more than just a technological advancement; it signifies a fundamental paradigm shift in scientific inquiry. This transformation implies that the tools and processes that underpin this shift, particularly research software, have become critical components of scientific integrity and overall progress. Consequently, the systematic study of these computational aspects becomes inherently relevant to understanding and improving science itself.

To address this evolving environment, metascience has emerged as a critical field dedicated to empirical study and the general improvement of scientific research~\cite{1}. Often referred to as the ``science of science,'' metascience investigates how research is conducted, funded, and evaluated, with the overarching goal of improving the efficiency, rigor, and overall impact of scientific research~\cite{1}. The contemporary relevance of metascience is underscored by significant institutional commitments, such as the recent formation of the Metascience Alliance, supported by the Center for Open Science, which aims to foster a global community dedicated to improving research practices. The rise and growing importance of metascience is directly related to systemic issues within scientific practice, such as the ``reproducibility crisis''~\cite{4}. This connection suggests a reactive, yet proactive effort to address fundamental flaws, highlighting metascience's role as a corrective and improvement-oriented discipline. The need for metascience arose from observed problems like irreproducibility and biases, and this problem-solving orientation is a key characteristic of an applied science, even when applied to science itself.

Scientific discovery today also has a unique need compared to the past, being the essential requirement for specialized research software~\cite{1}. In response to this, Research Software Science (RSS) is a field that studies how research software is used and developed by applying the principles of the scientific method~\cite{1, 7}. This paper aims to examine whether Research Software Science qualifies as a metascience by thoroughly exploring the definitional scope and characteristics of both fields, analyzing arguments for and against this classification, and synthesizing findings to offer a nuanced conclusion.

\section{Defining Metascience}

Metascience is broadly defined as the scientific study of science itself, with the explicit aim to describe, explain, evaluate, and ultimately improve scientific practices~\cite{4}. It delves into the methodology and philosophical implications of scientific investigation~\cite{4}. From an epistemological perspective, metascience builds conceptual knowledge about science by studying ``scientific constructs'' rather than the concrete objects studied by primary sciences. This distinction is crucial: metascience investigates the process and products of scientific knowledge creation, not the natural world directly~\cite{9}. This core distinction highlights that metascience studies the \textbf{how} and \textbf{what} of scientific knowledge itself---its constructs and processes. Any field that examines the mechanisms of science, rather than just a specific domain, could potentially be a metascience, which sets a rigorous standard for RSS to meet.

The overarching goal of metascience is to improve the efficiency, rigor, and overall impact of scientific research~\cite{1}. This includes addressing systemic problems within the research enterprise, realigning research assessments, and fostering research integrity~\cite{5}. Metascience interrogates every stage of the research lifecycle, from conception to publication and dissemination~\cite{5}. Its purpose is to make science better and enable scientists to conduct more robust research~\cite{4}. The ambition of metascience to ``overcome systemic problems of the research enterprise''~\cite{5} indicates its focus on the health and efficacy of the scientific system as a whole, extending beyond individual research projects. This system-level intervention is a defining characteristic of a meta-discipline, distinguishing it from efforts to merely improve individual research practices.

Metascience can be categorized into five major areas of interest: Methods, Reporting, Reproducibility, Evaluation, and Incentives. These areas correspond to how research is performed, communicated, verified, evaluated, and rewarded~\cite{6}. Key principles include: studying scientific practices, identifying poor research practices, biases, and misuse of statistics~\cite{6}; promoting transparency by advocating for pre-registration of studies, immediate data availability, and standardized reporting guidelines (e.g., CONSORT, EQUATOR Network)~\cite{4}; addressing the reproducibility crisis by investigating why many studies are difficult or impossible to replicate, especially in fields like psychology and medicine, as reproducibility and research integrity are essential tenets of scientific discovery~\cite{5}; evaluating peer review and funding by creating a scientific foundation for peer review and developing better research funding criteria~\cite{6}; scientific data science, which involves the use of data science to analyze research papers, encompassing fraud detection and citation network analysis~\cite{20}; and journalology, the scholarly study of all aspects of the academic publishing process, aiming to improve quality through evidence-based practices~\cite{6}.

The explicit mention of ``Scientific data science'' and ``Addressing bias in big data and AI''~\cite{5} within metascience's scope directly links it to computational methods and software. This indicates that metascience already recognizes the study of computational aspects of science as part of its domain. If metascience already includes studying how data science is applied to research papers and how artificial intelligence introduces bias in scientific processes, then the study of research software naturally falls within its purview.

To encompass the interdisciplinary nature of metascience, the general definition of scientific study of science itself, aiming to describe, explain, evaluate, and improve scientific practices~\cite{8}. Its core goals are to improve the efficiency, rigor, and overall impact of scientific research, making science better and enabling more robust research~\cite{4}. Key principles and areas include Methods, Reporting, Reproducibility, Evaluation, and Incentives, as well as studying scientific practices (e.g., biases, statistics misuse) and promoting statistical honesty~\cite{4}. From a philosophical perspective, metascience builds conceptual knowledge about science through the study of scientific constructs and epistemic operations (how knowledge is acquired, created, validated)~\cite{9}. Its goals include uncovering the true nature of reality through adequate interpretation of scientific knowledge and accounting for the relationship between science and society, focusing on metascientific epistemology, ontology, and semantics, and scientific constructs rather than natural objects~\cite{9}. Practically and problem-orientedly, metascience is an inquiry into the methodology and philosophical implications of scientific investigation, addressing issues like the reproducibility crisis~\cite{4}. Its goals are to overcome systemic problems of the research enterprise, realign research assessments, regain trust in science, and direct resources into effective research~\cite{4}. This includes addressing the reproducibility crisis, promoting transparency (pre-registration, data availability), evaluating peer review and funding, scientific data science, and journalology~\cite{4}.

\section{Defining Research Software Science (RSS)}

Research Software Science (RSS) is defined as applying the scientific method to understanding and improving how software is developed and used for research~\cite{7}. It promotes the use of scientific methodologies to explore and establish broadly applicable knowledge regarding research software~\cite{7}. RSS involves formal observation and experimentation to obtain knowledge through repeatable and reproducible processes. Obtaining data to detect correlations, designing experiments to identify cause and effect, and publishing results creating a broad impact is paramount to RSS~\cite{7}.

The primary objective of RSS is to pursue sustainable, repeatable, and reproducible software improvements that positively impact research software toward improved scientific discovery~\cite{7}. It aims to make software development and use in research more effective~\cite{7}. RSS seeks to elevate the status of software engineering to a scientific discipline within the research context, attracting talent and increasing respect for software work at research institutions~\cite{7}. The explicit aim of RSS to ``facilitate direct funding for software efforts''~\cite{7} and ``elevate respect for software efforts''~\cite{7} suggests that its scope extends beyond merely improving software. It also seeks to improve the recognition and resourcing of software development within the scientific ecosystem, aligning with metascience's concern for the overall health of the scientific enterprise. If RSS's goals extend to influencing funding and recognition, it operates at a level that impacts the system of science, not just the technical output.

The pursuit of understanding and improving research software has strong technical, social, and cognitive components~\cite{7}. The technical component focuses on the purpose of research software in modeling and simulation of scientific theories, and the gathering, analysis, and understanding of scientific data. This requires deep domain knowledge, such as mathematics, physics, and engineering for computational fluid dynamics (CFD) software~\cite{7}. The social component recognizes that scientific software development is typically a team effort, increasingly involving diverse roles, and emphasizes that team interactions, workflows, and tools significantly impact effectiveness~\cite{7}. Applying a scientific approach to studying and improving these social elements is crucial~\cite{7}. The cognitive component focuses on the learning processes involved in improving approaches to developing and using research software~\cite{7}. This includes leveraging knowledge from cognitive sciences to understand how developers and users approach their work and interact, and framing change as a problem or puzzle to engage scientists in the creative process~\cite{7}.

The emphasis on social and cognitive components reflects the increasing complexity and interdisciplinary nature of modern scientific software development~\cite{7}. This evolution necessitates a scientific approach that goes beyond traditional software engineering, aligning with metascience's holistic view of scientific practice. As scientific software becomes more complex and involves diverse teams, the human and organizational factors become as critical as the technical ones. Studying these factors scientifically, as RSS proposes, means RSS investigates the \textbf{human element} of science, which is a key area for metascience.

RSS is proposed as a complementary approach to Research Software Engineering (RSE) efforts for improving how software is developed and used for research~\cite{7}. While RSEs focus on practical application and improvement, RSS provides a scientific foundation for generating broadly applicable knowledge~\cite{7}. A key distinction lies in their primary aims: ``A scientist builds in order to learn; an engineer learns in order to build''~\cite{19}. RSEs seek improved tools or processes, test possibilities, and select the best, often with incidental team memory and limited focus on dissemination of generalizable knowledge~\cite{7}. RSS practitioners, conversely, aim to understand underlying principles, correlation, and cause-and-effect, designing studies, capturing data, analyzing results, and publishing findings~\cite{7}. This differentiation highlights an epistemological distinction: RSS is positioned as a knowledge-generating discipline about software development in research, whereas RSE is an applied discipline focused on direct implementation. This distinction strengthens the argument for RSS's scientific, and potentially meta-scientific, nature. If RSS is about \textbf{generating knowledge} about how research software is developed and used, rather than just applying existing knowledge to build software, it aligns more closely with the foundational, inquiry-driven nature of science, and by extension, metascience.

\section{Arguments for Research Software Science as a Metascience}

\subsection{Shared Approach to Studying Scientific Processes}

Both metascience and RSS look to advance scientific knowledge through empirically studying processes~\cite{1}. Metascience uses empirical methods to investigate how research is conducted, funded, and evaluated~\cite{1}. Similarly, RSS applies the scientific method---involving formal observation and experimentation---to understand and improve software development and use in research~\cite{7}. This shared commitment to empirical methodologies (observation, experimentation, data analysis) represents a fundamental thematic link. Both fields apply scientific rigor to a process (science itself for metascience, software development for RSS), rather than just using scientific methods to study natural phenomena. This methodological parallelism suggests that RSS operates on a similar meta-level, applying scientific methods to a process that underpins scientific discovery.

\subsection{Overlapping Principles: Reproducibility, Transparency, and Rigor}

The overlapping principles of core concepts show a deep connection between RSS and metascience~\cite{1}.

\textbf{Reproducibility}: Metascience identifies the replication crisis as a major concern and seeks solutions~\cite{6}. RSS directly contributes to computational reproducibility, which is a simple premise in theory but often difficult to achieve in practice~\cite{2}. RSS promotes practices such as literate programming, robust code version control and sharing, effective compute environment control (e.g., through containers), persistent data sharing, and comprehensive documentation to ensure computational research work can be reproduced quickly and easily~\cite{2}. These practices directly support metascience's goals of enhancing research rigor and reliability~\cite{2}. RSS's direct focus on improving reproducibility and transparency in computational research means it is actively \textbf{enabling and implementing} core goals of metascience. RSS is not merely a tool; it is a domain of scientific inquiry that provides the empirical basis and practical solutions for meta-scientific problems in computational contexts. If metascience identifies problems, such as the reproducibility crisis, and seeks solutions through empirical study, and RSS provides empirical solutions to these problems within the computational domain, then RSS is effectively a specialized branch or application of metascience.

\textbf{Transparency}: Research transparency is a major aspect of scientific research, covering data and code sharing, citation standards, and verifiability~\cite{3}. Metascience advocates for greater transparency in scientific studies, including better requirements for disclosure and standardized reporting~\cite{6}. RSS, through its focus on open science principles and adherence to guidelines like the Transparency and Openness Promotion (TOP) guidelines for data and analytic methods transparency, directly supports these metascience objectives~\cite{3}. Transparency also involves functional transparency (knowledge of algorithmic functioning) and structural transparency (knowing how a program instantiates an algorithm).

\subsection{Alignment with Broader Goals of Scientific Discovery}

From a larger scale perspective, RSS fundamentally is a process that helps further scientific knowledge through research, which aligns with broad definitions of metascience~\cite{1}. RSS aims for sustainable, repeatable, and reproducible software improvements that positively impact research software toward improved scientific discovery~\cite{7}. The goals of RSS require collaboration with cognitive and social scientists, accessing their expertise and scientific processes and tools~\cite{7}. This interdisciplinary nature, studying the human and organizational aspects of scientific software, mirrors metascience's holistic examination of the scientific enterprise.

\section{Arguments Against Research Software Science as a Metascience}

\subsection{RSS as a Tool within Science vs. the ``Science of Science''}

A primary argument against RSS being a metascience is that it is defined as applying the scientific method to understanding and improving the development and use of research software~\cite{7}. Following this definition, RSS could be considered a tool within science, rather than the ``science of science'' itself~\cite{1}. Scientific disciplines are generally distinguished by the nature of the phenomena of interest, the kinds of questions asked, and the types of tools, methods, and techniques used~\cite{16}. If RSS's ``phenomenon of interest'' is primarily the software artifact, it positions RSS as a specialized scientific discipline (e.g., like materials science studies materials, or computer science studies computation) rather than a discipline studying the broader scientific enterprise. The fundamental counter-argument hinges on whether RSS's primary object of study is \textbf{software} (a tool or artifact) or the \textbf{scientific process itself}~\cite{1}. If it is the former, it is a science that \textbf{uses} scientific methods to study a specific domain (software). If it is the latter, it is a meta-science. This distinction represents the crux of the debate. Science studies the natural world through empirical means~\cite{17}. Software, while crucial for science, is a human construct. If RSS primarily studies the creation and properties of this construct, it is a science about software. Metascience, however, studies the processes and structures of science itself. The distinction lies in the primary object of inquiry.

\subsection{Distinction between Scientific Tools/Disciplines and the ``Science of Science''}

While individual scientific disciplines employ specialized tools and methods suited to their subject matter~\cite{16}, the field of metascience — the “science of science” — extends beyond disciplinary tools to analyze the fundamental principles, methodologies, and societal impacts that shape scientific practice across fields~\cite{8}. Research Software Science (RSS), by contrast, emphasizes the development and refinement of software tools and practices that support scientific work. However, unlike metascience, RSS does not primarily investigate the systemic processes by which scientific knowledge is generated, validated, or disseminated. Instead, RSS addresses the means (improving software tools and workflows) that enable scientific research, whereas metascience examines the ends — the broader effectiveness, integrity, and trustworthiness of the scientific enterprise itself. Therefore, unless RSS explicitly includes a critical, systemic perspective on how software infrastructure shapes the production and credibility of scientific knowledge, it may not fully reach the “meta” level of inquiry that defines metascience.

\subsection{Positioning RSS as a Specialized Domain within Science}

Given its definition, RSS could be more accurately positioned as a specialized domain within computer science or software engineering that applies scientific rigor to a specific type of software (research software)~\cite{1}. This would make it a scientific discipline in its own right, but not necessarily a ``meta'' science that operates above or across all scientific fields. Early scientific teams evolved from domain experts to include mathematics, computer science, and software engineering~\cite{7}. RSS, by adding cognitive and social scientists to this evolution, represents a further specialization within the scientific team structure, aiming to optimize a particular aspect (software) of scientific endeavor~\cite{7}. This continuous refinement of specialized roles and disciplines is characteristic of scientific progress, not necessarily the emergence of a meta-discipline.

\section{Conclusion}

Whether Research Software Science (RSS) should be classified as a metascience remains a nuanced question, largely shaped by how metascience itself is defined~\cite{1}. Proponents of this classification point to RSS’s empirical approach~\cite{1}, its contributions to core meta-scientific goals such as reproducibility and transparency in computational research~\cite{1}, and parallels with Empirical Software Engineering. The rapid integration of artificial intelligence in software development~\cite{18} and emerging “metascience of software” studies further suggest that investigating software’s impact on science aligns with meta-scientific aims.

On the other hand, the argument that RSS primarily addresses \textbf{software as a tool} rather than the broader systems and epistemic structures that metascience typically examines~\cite{1}. The core distinction is whether RSS studies the \textbf{means} that support scientific work or the \textbf{system} that underpins scientific knowledge as a whole.

RSS applies rigorous methodologies to a critical enabler of modern science: research software. If metascience is defined broadly as any empirical effort that improves scientific practice and outcomes~\cite{8}, then RSS fits within this scope by strengthening reproducibility, transparency, and the social and cognitive dimensions of software development~\cite{2}. However, if metascience is narrowly defined as the study of fundamental epistemological frameworks and systemic structures of science itself~\cite{9}, RSS may be better described as an interdisciplinary field that supports metascience rather than being metascience in its own right.

Ultimately, RSS sits at the intersection of discipline-specific research and meta-scientific inquiry. It addresses how robust software sustains credible science. Recognizing RSS as a distinct scientific domain — whether as a branch of metascience or a closely allied interdisciplinary science — is essential for its continued development. This recognition can justify dedicated funding~\cite{7} and elevate the status of software work within research institutions~\cite{7}. It also reinforces the application of scientific rigor to the development of research software, ensuring that the tools of discovery meet the same standards as the discoveries themselves. The growing involvement of cognitive and social scientists in RSS highlights its expanding role in strengthening both the technical and human aspects of computational research~\cite{7}.

\printbibliography

@online{1,
  author = {{Center for Open Science}},
  title = {{Metascience Alliance}},
  url = {https://www.cos.io/metascience-alliance},
  urldate = {2025-07-16},
  organization = {Center for Open Science},
}

@article{2,
    author = {Ziemann, Mark and Poulain, Pierre and Bora, Anusuiya},
    title = {The five pillars of computational reproducibility: bioinformatics and beyond},
    journal = {Briefings in Bioinformatics},
    volume = {24},
    number = {6},
    pages = {bbad375},
    year = {2023},
    month = {10},
    issn = {1477-4054},
    doi = {10.1093/bib/bbad375},
    url = {https://doi.org/10.1093/bib/bbad375},
    eprint = {https://academic.oup.com/bib/article-pdf/24/6/bbad375/52346129/bbad375.pdf},
}

@article{3,
  author    = {Hans Ekkehard Plesser},
  title     = {Reproducibility vs. Replicability: A Brief History of a Confused Terminology},
  journal   = {Frontiers in Neuroinformatics},
  volume    = {11},
  pages     = {76},
  year      = {2018},
  month     = {1},
  doi       = {10.3389/fninf.2017.00076},
  pmid      = {29403370},
  pmcid     = {PMC5778115}
}

@online{4,
  title = {{The Rise of Metascience | Metascience - BioTechniques}},
  url = {https://www.biotechniques.com/technology-news/the-rise-of-metascience/},
  urldate = {2025-07-20},
}

@book{5,
  author    = {Bennett, Gavin},
  title     = {An Introduction to Metascience: The Discipline of Evaluating the Creation and Dissemination of Research},
  year      = {2024},
  publisher = {Taylor and Francis},
  address   = {Abingdon, UK},
  
}

@article{6,
  author    = {Ioannidis, John P. A. and Fanelli, Daniele and Dunne, Desmond D. and Goodman, Steven N.},
  title     = {Meta-research: Evaluation and Improvement of Research Methods and Practices},
  journal   = {PLoS Biology},
  year      = {2015},
  month     = {10},
  volume    = {13},
  number    = {10},
  pages     = {e1002264},
  doi       = {10.1371/journal.pbio.1002264},
  pmid      = {26431313},
  pmcid     = {PMC4592065}
}

@article{7,
  author = {Michael A. Heroux},
  title = {{Research Software Science: Expanding the Impact of Research Software Engineering}},
  year      = {2022},
  journal = {DigitalCommons@CSBSJU}, 
}

@online{8,
  author = {{FORRT - Framework for Open and Reproducible Research Training}},
  title = {{Meta-science or Meta-research | FORRT}},
  url = {https://forrt.org/glossary/vbeta/meta-science-or-meta-research/},
  urldate = {2025-07-20},
}

@article{9,
	author = {Fran\c{c}ois Maurice},
	journal = {M\ensuremath{\epsilon}tascience: Scientific General Discourse},
	pages = {1--312},
	title = {M\ensuremath{\epsilon}tascience: Scientific General Discourse - No. 3 - Metascientific Epistemology},
	volume = {3},
	year = {2024}
}

@book{16,
  author = {{National Academies of Sciences, Engineering, and Medicine}},
  title = {{Scientific Methods and Knowledge - Reproducibility and ...}},
  publisher = {National Academies Press},
  url = {https://www.ncbi.nlm.nih.gov/books/NBK547541/},
  urldate = {2025-07-20},
  year = {2019}, 
}

@online{17,
  title = {{Science as a Way of Knowing - Climate Science Investigations ...}},
  url = {https://www.ces.fau.edu/nasa/introduction/science-way-knowing.php},
  urldate = {2025-07-20},
  organization = {Florida Atlantic University},
}

@article{18,
  author = {{ResearchGate}},
  title = {{The Evolution of Software Engineering: Navigating the AI-Driven Development Landscape}},
  journal = {ResearchGate},
  url = {https://www.researchgate.net/publication/390208939_The_Evolution_of_Software_Engineering_Navigating_the_AI-Driven_Development_Landscape},
  urldate = {2025-07-20},
  year = {2024},
}

@article{19,
author = {Brooks, Frederick P.},
title = {The computer scientist as toolsmith II},
year = {1996},
issue_date = {March 1996},
publisher = {Association for Computing Machinery},
address = {New York, NY, USA},
volume = {39},
number = {3},
issn = {0001-0782},
url = {https://doi.org/10.1145/227234.227243},
doi = {10.1145/227234.227243},
journal = {Commun. ACM},
month = mar,
pages = {61–68},
numpages = {8}
}

@article{20,
  author    = {Markowitz, David M. and Hancock, Jeffrey T.},
  title     = {Linguistic Obfuscation in Fraudulent Science},
  journal   = {Journal of Language and Social Psychology},
  year      = {2015},
  volume    = {35},
  number    = {4},
  pages     = {435--445},
  doi       = {10.1177/0261927X15614605},
  note      = {Original work published 2016},
  url       = {https://doi.org/10.1177/0261927X15614605}
}

\end{document}